\documentclass[noshowpacs,amsmath,twocolumn,superscriptaddress,10pt,aps,prb]{revtex4-1} 
\usepackage{setspace}
\usepackage{amsmath}
\usepackage{breqn}
\usepackage{graphicx}
\usepackage[nearskip,margin = 0pt,caption=false]{subfig}

\usepackage{verbatim}
\usepackage{amsfonts}
\usepackage{amssymb}
\usepackage{epstopdf}
\usepackage{lipsum}
\usepackage{siunitx}
\usepackage{xcolor}
\usepackage{array}
\usepackage[normalem]{ulem}
\usepackage[resetlabels,labeled]{multibib}
\DeclareGraphicsExtensions{.pdf,.eps,.png,.jpg,.mps}
\usepackage{etoolbox}

\begin{document}
\title{Inverse-designed multi-dimensional silicon photonic transmitters}
\author{K.Y.Yang$^{1,*}$, A.D.White$^{1,*}$, F.Ashtiani$^{2,*}$, C.Shirpurkar$^{3,*}$, 
S.V.Pericherla$^{3,*}$, L.Chang$^{4}$, H.Song$^{5}$, K.Zou$^{5}$, H.Zhou$^{5}$, K.Pang$^{5}$, J.Yang$^{1}$, M.A.Guidry$^{1}$, D.M.Lukin$^{1}$, H.Hao$^{2}$, L.Trask$^{3}$, G.H.Ahn$^{1}$, A.Netherton$^{4}$, T.C.Briles$^{6}$, J.R.Stone$^{6}$, L.Rechtman$^{7}$, J.S.Stone$^{8}$, K.Van Gasse$^{1}$, J.L.Skarda$^{1}$, L.Su$^{1}$, D.Vercruysse$^{1}$, J.P.W.MacLean$^{1}$, S.Aghaeimeibodi$^{1}$, M.-J.Li$^{8}$, D.A.B.Miller$^{1}$, D.Marom$^{7}$, S.B.Papp$^{6,9}$, A.E.Willner$^{5}$, J.E.Bowers$^{4}$, P.J.Delfyett$^{3}$, F.Aflatouni$^{2}$, \\and J.Vu\v{c}kovi\'{c}$^{1,\dagger}$\\
\vspace{+0.05 in}
$^1$E.L.Ginzton Laboratory, Stanford University, Stanford, CA, USA.\\
$^2$Department of Electrical and Systems Engineering, University of Pennsylvania, Philadelphia, PA, USA.\\
$^3$The College of Optics and Photonics, University of Central Florida, Orlando, FL 32816, USA.\\
$^4$Department of Electrical and Computer Engineering, University of California, Santa Barbara, CA, USA.\\
$^5$Department of Electrical Engineering, University of Southern California, Los Angeles, CA, USA\\
$^6$Time and Frequency Division, National Institute of Standards and Technology, Boulder, CO, USA\\
$^7$Applied Physics Department, The Hebrew University of Jerusalem, Jerusalem, Israel\\
$^8$Corning Incorporated, Corning, NY, USA\\
$^9$Department of Physics, University of Colorado, Boulder, CO, USA\\
{\small $^{\dagger}$Corresponding author: jela@stanford.edu, *These authors contributed equally to this work.}}

\begin{abstract}
    \noindent Modern microelectronic processors have migrated towards parallel computing architectures with many-core processors. However, such expansion comes with diminishing returns exacted by the high cost of data movement between individual processors. The use of optical interconnects\cite{Miller:2009:IEEE,Miller:2017:JLT} has burgeoned as a promising technology that can address the limits of this data transfer. While recent pushes to enhance optical communication have focused on developing wavelength-division multiplexing technology, this approach will eventually saturate the usable bandwidth, and new dimensions of data transfer will be paramount to fulfill the ever-growing need for speed\cite{Willner:2012:NaturePhotonics,Siddharth:2013:Science,Richardson:2013:NaturePhotonics,Kahn:2017:NaturePhotonics}. Here we demonstrate an integrated intra- and inter-chip multi-dimensional communication scheme enabled by photonic inverse design. Using inverse-designed mode-division multiplexers, we combine wavelength- and mode- multiplexing and send massively parallel data through nano-photonic waveguides and optical fibres. Crucially, as we take advantage of an orthogonal optical basis, our approach is inherently scalable to a multiplicative enhancement over the current state of the art.
\end{abstract}

\maketitle

\begin{figure*}[t!]
\centering
\includegraphics[width=0.925\linewidth]{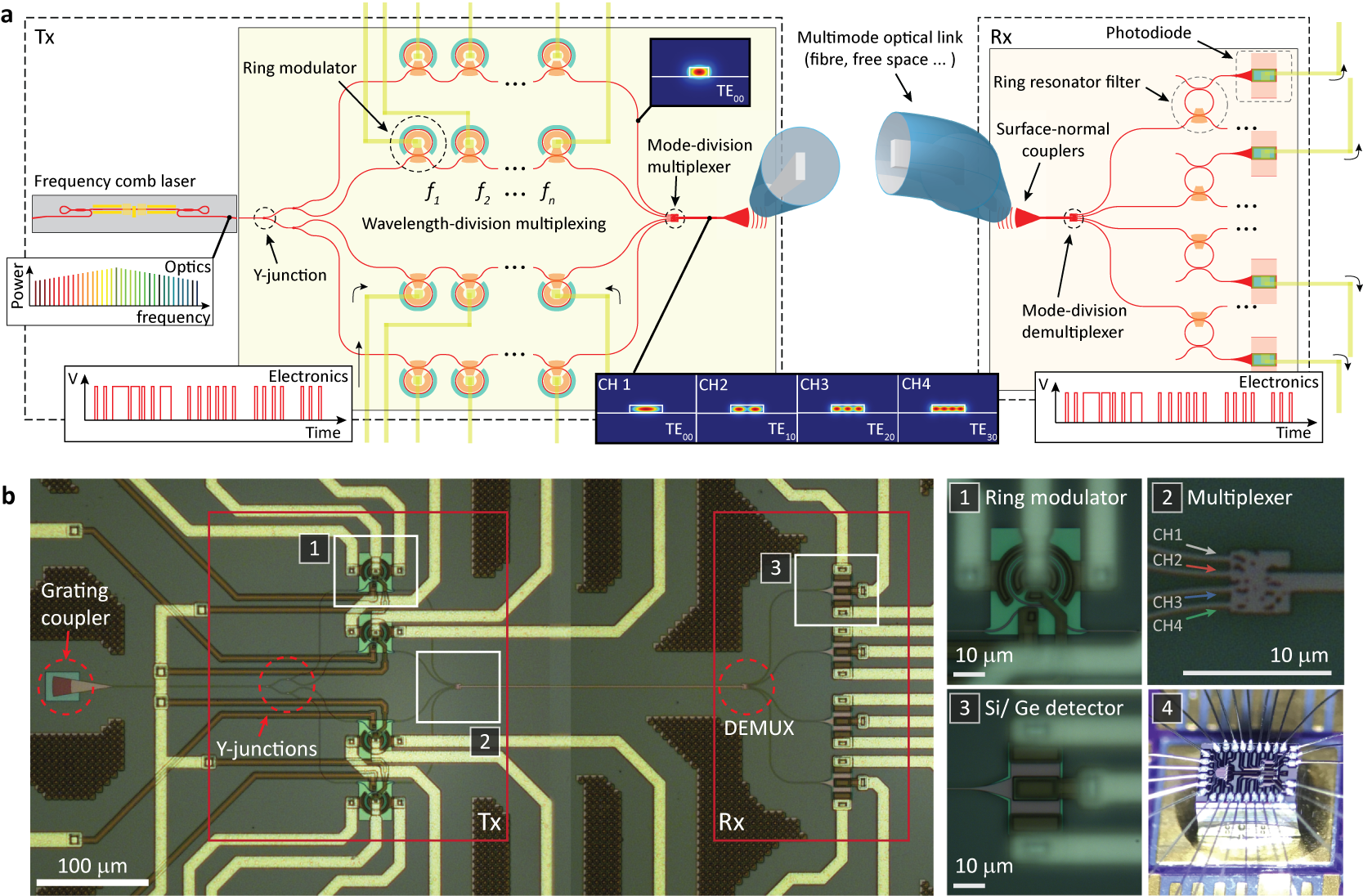}
\caption{\label{fig:Fig1}{\bf{An optical transmitter using multi-dimensional multiplexing}} (\textbf{a}) Principle of chip-to-chip WDM-MDM data transmission. A frequency comb laser is evenly distributed to individual WDM transmitters (ring modulators) using cascaded Y-junction splitters. The multi-frequency data from each WDM transmitter are routed into spatial modes of a multi-mode waveguide using an MDM multiplexer. The optical data can then be transmitted through a multimode optical link to the receiver, where the mode and wavelength channels are separated by MDM-WDM demultiplexers and detected using photodiodes. (\textbf{b}) Microscope image of integrated four channel MDM transceiver (Inset 1: ring modulator, Inset 2: MDM multiplexer, Inset 3: Si/Ge photodetector, Inset 4: wire-bonded chip).}
\end{figure*}


\begin{figure*}[t!]
\centering
\includegraphics[width=0.90\linewidth]{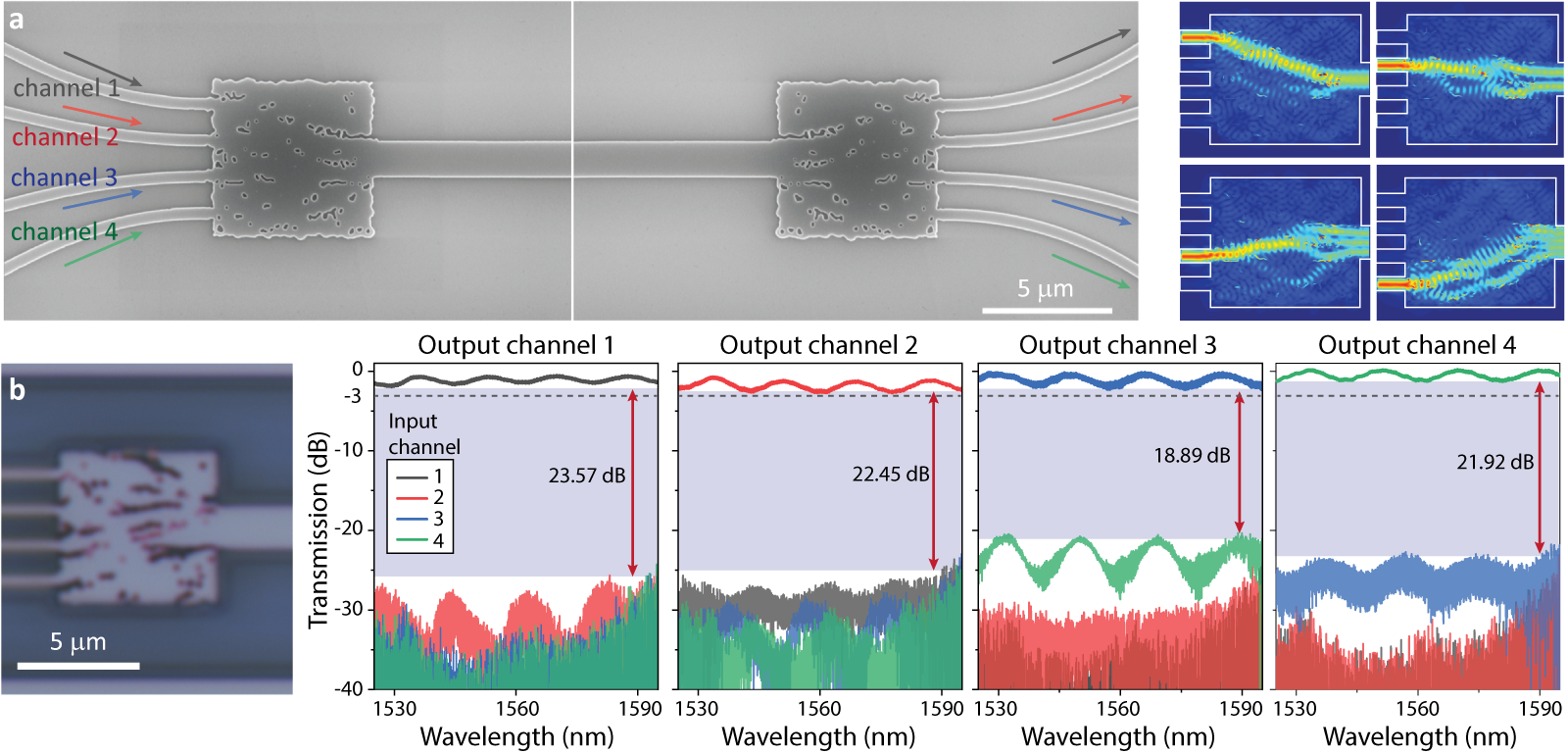}
\caption{\label{fig:Fig2}{\bf{Photonic inverse design of broadband, low-crosstalk, all-passive MDM multiplexers.}} \textbf{(a)} SEM images of back-to-back MDM multiplexer devices with simulated mode conversion (right). (\textbf{b}) Measured S-parameters of the back-to-back MDM multiplexer structure. The structure is optimized at multiple wavelengths over the entire C band. The broadband structure is utilized in the WDM-MDM data transmission experiments using soliton microcombs (13-wavelength channels, 300 GHz channel spacing over 3.6 THz spectral span; see Fig.\ref{fig:Fig3}).}
\end{figure*}

\noindent Rapid advances in computing and networking have led to an explosion of new technology in areas ranging from genomics\cite{Alipanahi:2015:NatureBioTech} and medicine\cite{Topol:2019:NatureMedicine} to multimedia streaming to artificial intelligence\cite{Weiss:1999,Haddadi:2019:IEEE,Subhasish:2021:NatureElectronics}. Now, as computation has become increasingly parallelised and offloaded to the cloud\cite{Berkeley:2010,IBM:2017}, the limits of performance lie in the interconnects between computational nodes. In data centers, massive amounts of information need to be constantly transferred between many individual processors and memory. While this can be done electrically, the higher frequency and lower loss of on-chip electro-optical interconnects allow for wider bandwidth and greater energy efficiency\cite{Miller:2009:IEEE,Miller:2017:JLT}. Optical interconnects are further enhanced by wavelength-division multiplexing (WDM), which enables parallel signal transmission by independently encoding data on multiple frequencies of light\cite{Stojanovic:2015:Nature,Koos:2017:Nature,Bowers:2019:Optica,Ballani:2020}. However, this approach is still fundamentally limited to the bandwidth that optical fibres can support. To further increase the bandwidth of links, other dimensions of signal encoding must be used.

One promising dimension that can be utilized for multiplexing is the spatial domain; light can be decomposed into a set of optical beams with orthogonal spatial cross-sections. These orthogonal spatial modes in multi-mode optical waveguides or in free space can serve as independent communication channels\cite{Miller:2019:Tutorial,Willner:2012:NaturePhotonics,Richardson:2013:NaturePhotonics,Siddharth:2013:Science,Ryf:2011:JLT,Van:2014:NaturePhotonics,Kahn:2017:NaturePhotonics,Puttnam:2020:JLT,Rademacher:2020:ECOC}, each of which can support a full WDM link. This independence gives mode-division multiplexing (MDM) a multiplicative effect on the bandwidth of an optical link. Recently, significant progress has been made towards integrating mode and wavelength division multiplexing together on chip\cite{Gabrielli:2012:NatureCommunications,Luo:2014:NatureCommunications,Miller:2020:Optica,Dai:2018:LPR,Baumann:2018:OFC,Bell:2020}.

In this work, we present a multi-wavelength, multi-mode communication scheme for on-chip and chip-to-chip interconnects. Using photonic inverse design\cite{Piggott:2015:NaturePhotonics,Molesky:2018:NaturePhotonics}, we implement a low-crosstalk, all-passive silicon MDM that supports parallel WDM channels over a 15-THz spectral bandwidth. With this device, we demonstrate error-free parallel data transmission using an integrated frequency comb laser\cite{Papp:2020:PhyRevAppl} and multi-dimensional multiplexing: 52 data channels derived from 13 wavelength channels launched into four spatial mode channels. The MDM device also enables spectrally efficient WDM due to its non-resonant operation, allowing us to demonstrate a 1.28-Tb/s aggregated data rate using only 320 GHz of the available spectral bandwidth. Additionally, in combination with inverse designed beam emitters and a multimode-matched fibre\cite{Marom:2019:JSTQE,Marom:2017:IPC}, we demonstrate chip-to-chip data transmission across four spatial mode channels at three different wavelengths between separate silicon chips. \\

\noindent\textbf{A silicon multi-dimensional transmitter}  

\noindent We use inverse design to implement broadband multi-dimensional multiplexing on an integrated silicon photonics platform. We show that inverse-designed MDM devices can operate over a large, uninterrupted spectral bandwidth, and are thus able to be combined with silicon photonic WDM circuits to achieve orthogonal data encoding. In addition, the multiplexed data can be mapped using silicon photonics to orthogonal spatial channels of free space or optical fibres for chip-to-chip data transmission. The principles of this massively parallel optical communications based on WDM-MDM are illustrated in Fig.\ref{fig:Fig1}a. A common multi-frequency laser source\cite{kallfass:2013:NaturePhotonics,Koos:2017:Nature,Oxenlowe:2018:NaturePhotonics,Loncar:2019:Nature,Moss:2020:NatureCommunications,Bowers:2020:Nature,Papp:2020:PhyRevAppl} is evenly distributed into multiple WDM transmitter circuits, and each WDM circuit independently encodes data onto different frequencies of light. An inverse-designed MDM multiplexer takes the overlapping modes from multiple WDM transmitters and transforms them into co-propagating spatially orthogonal modes of a multi-mode optical waveguide. An efficient chip-to-chip optical link (e.g. surface normal beam emitter) enables the transmission of the WDM-MDM data between separate silicon chips throughout free space or optical fibre. Finally, a receiver circuit de-multiplexes the spatially and spectrally mapped signals and converts the data to electronic circuits via photodetectors.

Fig.\ref{fig:Fig1}b shows a simplified implementation of the proposed communication scheme -- four MDM channel transceiver with one ring modulator per spatial mode (4-mode, 1-wavelength/mode). The chip was fabricated using a commercial silicon foundry process, and the devices were designed to comply with the process design rules\cite{Piggott:2015:NaturePhotonics,Su:2020,Piggott:2020:ACSPhotonics} (details of the design and process are described in the Methods). The integrated system consists of four ring modulators, MDM multiplexers, and Si/Ge photodetectors. The MDM multiplexer, optimized for operation in the communication wavelength band (Inset 2 of Fig.\ref{fig:Fig1}b), routes data from each ring modulator into orthogonal spatial modes (TE$_{00}$, TE$_{10}$, TE$_{20}$, TE$_{30}$). It is important to note that the MDM multiplexer doesn't require post-processing/trimming or temperature tuning. To verify the operation of the MDM link, we measured an open eye diagram of the integrated MDM transceiver and the modal cross-talk of the all-passive MDM structure which ranges from -10 to -20 dB (see Fig.\ref{fig:SI_IME_eye_diagram}). Our demonstration can be in principle scaled up to implement a massively parallel WDM-MDM data transmission in the next section.\\ 


\begin{figure*}[htbp]
\centering
\includegraphics[width=0.90\linewidth]{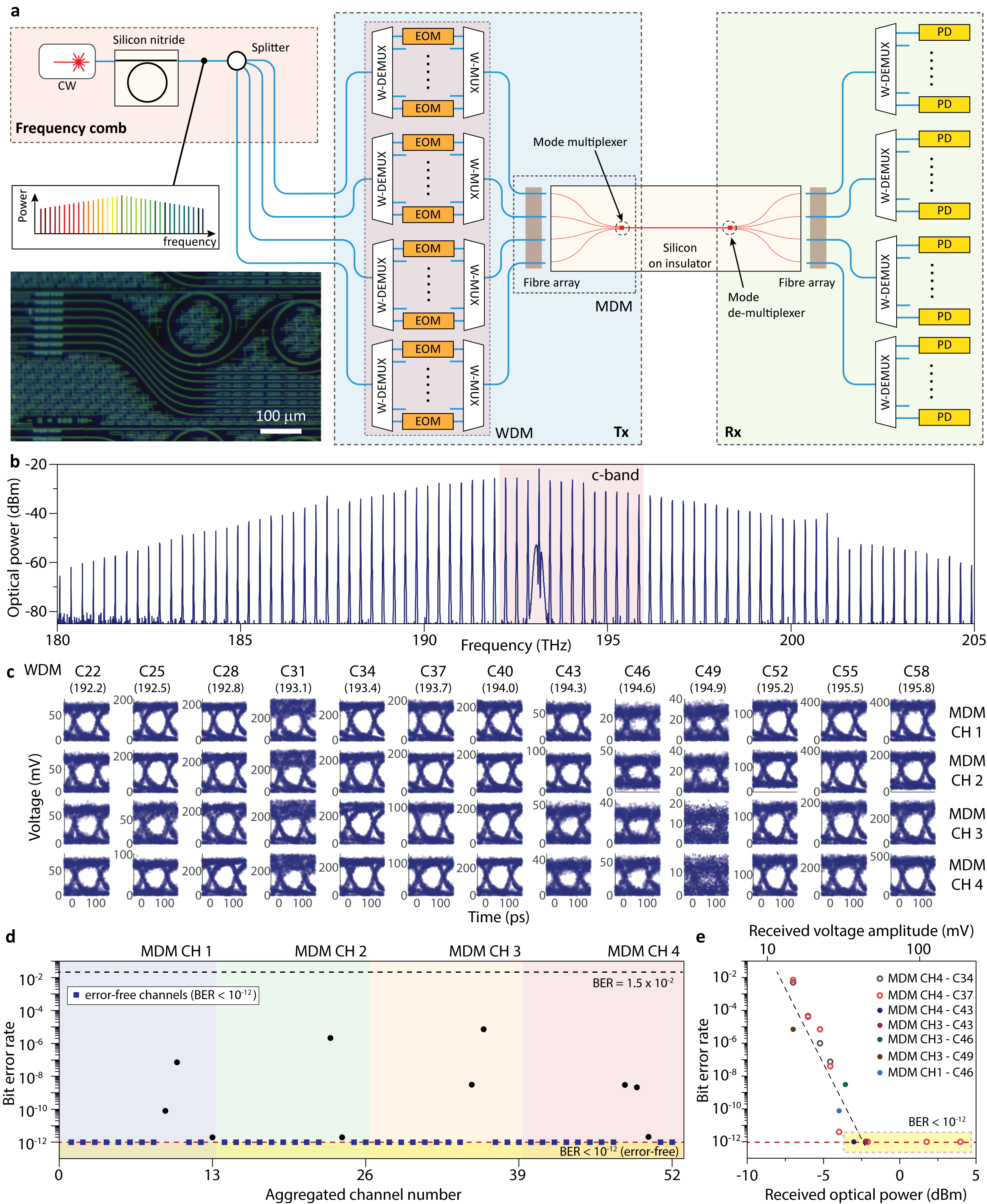}
\caption{\label{fig:Fig3}{\bf{On-chip interconnect: WDM - MDM data transmission with a soliton microcomb laser.}} \textbf{(a)} Parallel WDM-MDM data transmission scheme using soliton microcombs as a multi-wavelength source. The transmitter optical source is generated by pumping the Si$_{3}$N$_{4}$ microresonator (inset: optical microscope image of the device) with a CW laser. WDM de-multiplxers (W-DEMUX) separate the comb lines and intensity modulators (EOM) encode independent data (NRZ at a symbol rate of 10 GBd/channel). The WDM data are recombined using WDM multiplexers (W-MUX), and are coupled to on-chip MDM inputs simultaneously using a fibre array. At the receiver, all the channels are separated by mode and wavelength de-multiplexers and detected using a photodiode (PD). In our experiments, we independently modulate even and odd WDM channels using two EOMs for emulating WDM transmission (see Method section for more details). \textbf{(b)} Optical spectrum of the soliton microcomb. For the data transmission measurements, 13 comb lines at the C band (red shaded region) were modulated with 10-Gb/s NRZ signals. \textbf{(c)} 10-Gb/s eye diagrams of all data channels directly detected using PD. \textbf{(d)} Measured BERs (10$^{12}$ bits compared) of the transmitted data channels, and \textbf{(e)} BER sensitivity versus received optical power (or electrical voltage amplitude).}
\end{figure*}

\begin{figure*}[t!]
\centering
\includegraphics[width=0.90\linewidth]{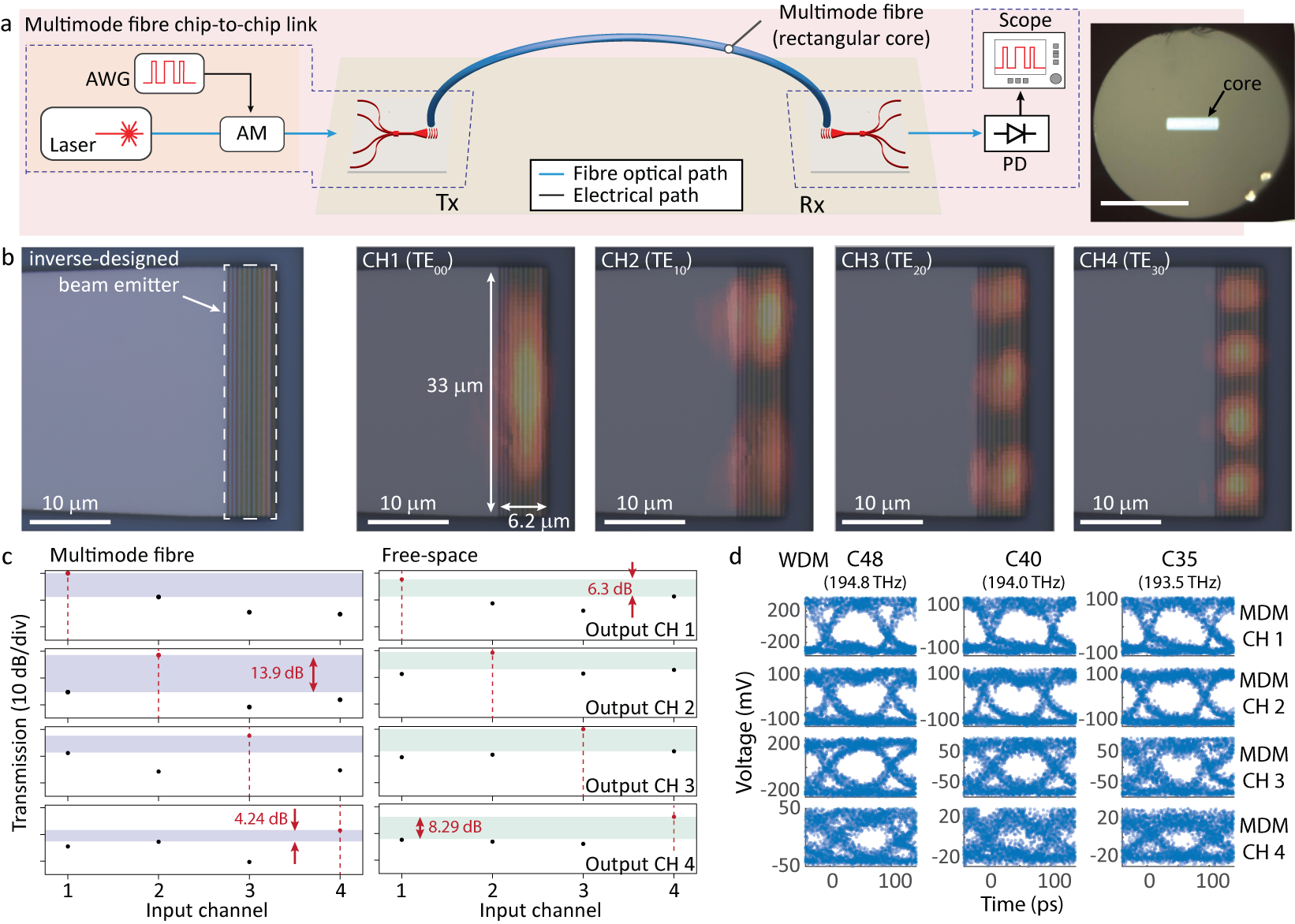}
\caption{\label{fig:Fig4}{\bf{Chip-to-chip interconnect using inverse-designed beam emitters and a rectangular core multimode fibre.}} \textbf{(a)} MDM chip-to-chip data transmission scheme using a multimode fibre. A CW laser is modulated using an amplitude modulator, and the data signal is coupled to a Tx input waveguide via a lensed fibre. The Tx chip sends multi-channel MDM beams to a multimode fibre using an inverse-designed coupler. The MDM data are sent to the Rx chip that separates each mode into a single-mode output. All data channels are directly detected using photodiodes at the receiver side. \textbf{(b)} Infrared images  of mode profiles at Tx-out measured at 1540 nm are overlaid on device microscope images (bar: 10 $\mu$m). \textbf{(c)} Measured chip-to-chip MDM channel cross-talks and insertion losses at 1540 nm (Left: multimode fibre chip-to-chip link, Right: free-space). \textbf{(d)} 10-Gb/s eye diagrams for a multimode fibre chip-to-chip link at three different frequencies. The signal degradation of the mode channel 4 is attributed to the mode-dependent insertion loss.} 
\end{figure*}

\noindent\textbf{On-chip interconnect}  

\noindent In this section, we show error-free data transmission using a multi-dimensional multiplexing and integrated laser sources -- demonstrating for the first time a silicon photonic MDM interconnect with a chip-scale frequency comb source. The MDM structure supports a broadband operation as well as an arbitrary channel spacing for high-speed dense WDM (DWDM) communications. We use external modulators and a 100-GHz-grid DWDM in this experiment; however the MDM technology is fully compatible with silicon photonic WDM transmitters\cite{Stojanovic:2015:Nature,Wade:2021:OFC,Bergman:2021:arXiv}. 

The inverse-designed MDM devices were fabricated in a standard silicon-on-insulator platform (silicon thickness: 220 nm). Fig.\ref{fig:Fig2}a shows a scanning electron micrograph (SEM) image of the MDM muliplexer, whose design area is 6.5 $\times$ 6.5 $\mu m^{2}$ with a foundry-compatible minimum feature size (80 nm)\cite{GF:2020:OFC}. The MDM multiplexer routes the fundamental transverse-electric mode (TE$_{00}$) of the 500-nm-wide input waveguides to the TE$_{00}$ (channel 1), TE$_{10}$ (channel 2), TE$_{20}$ (channel 3), and TE$_{30}$ (channel 4) modes of the 1800-nm-wide output waveguide. The device structure is optimized for operation over multiple wavelengths within the communication wavelength band, as broadband optimization has previously been shown to be an effective heuristic for fabrication robustness\cite{Piggott:2020:ACSPhotonics,Yang:2020:NaturePhotonics}. Fig.\ref{fig:Fig2}b shows measured S-parameters of the back-to-back MDM multiplexers. We find that the peak insertion loss is less than 0.8 dB and 3-dB bandwidth is wider than 150 nm for all mode channels. Additionally, the modal cross-talk, which must be suppressed to avoid MDM signal degradation \cite{Luo:2014:NatureCommunications}, is less than -18 dB for all mode channels. Notably, the cross-talk of the broadband design is significantly lower compared to that of the narrow band design (see Fig.\ref{fig:SI_MDM_design}), which can be attributed to fabrication-error robustness.

The conceptual diagram of WDM-MDM data transmission using Si$_{3}$N$_{4}$ soliton microcombs is depicted in Fig.\ref{fig:Fig3}a. The soliton microcombs are generated as circulating pulses in a chip-scale optical cavity pumped with a CW laser, leading to low-noise, broadband optical frequency combs\cite{Koos:2017:Nature,Papp:2020:PhyRevAppl}. In this experiment, a soliton source with 300-GHz frequency spacing (Fig.\ref{fig:Fig3}b) is the optical source of the transmitter and all data channels are directly detected using photodiodes. We use non-return-to-zero (NRZ) coding at a symbol rate of 10 GBd, and simultaneously launch 13 WDM channels into 4 MDM channels. It is important to note that we use optical fibre amplifiers (see Fig.\ref{fig:SI_soliton_WDM_MDM_setup}) to compensate for the insertion loss of external optical components (e.g. modulators, DWDM) but this extra loss can be mitigated once all components are integrated.

The eye diagrams and bit error rates (BERs) of the soliton data transmission experiment are presented in Fig.\ref{fig:Fig3}c-d. Out of the 52 carriers derived from the four spatial mode channels and all comb lines in the C band, 42 data channels show no error occurrences when a total of 10$^{12}$ bits are compared (BER $<$ 10$^{-12}$). The data channels also show BERs better than 10$^{-10}$ with 3 dBm input optical power at 10 Gb/s/channel (see Fig.\ref{fig:Fig3}e and Fig.\ref{fig:SI_soliton_WDM_MDM_setup}c). The total data rate in this experiment amounts to 520 Gb/s. Additionally, we show that the optical link can also be driven at 20 Gb/s/channel with open eye diagrams to achieve an aggregate nano-photonic waveguide bandwidth of 1.04 Tb/s (see Fig.\ref{fig:SI_soliton_WDM_MDM_20gbps_eye_diagram}). In this experiment, data channels at specific WDM channels (e.g. C46 and C49) show larger error rates than other wavelength channels, owing to lower comb line power at the WDM channel and non-uniformity of the soliton optical spectrum (see Fig.\ref{fig:SI_soliton_WDM_MDM_setup}b).

In a single MDM device, which can achieve a spectral bandwidth of over 150 nm (see Fig.\ref{fig:Fig2}), further improvement in the data transmission capacity is straightforward through the use of adjacent S and L bands. We can also decrease the frequency spacing of the comb source to the standard 100-GHz frequency grid\cite{Koos:2017:Nature} or even finer\cite{Bowers:2020:Nature}. As a proof of concept, we demonstrate WDM-MDM data transmission (4-mode, 16-wavelength/mode) using a mode-locked laser with a 20-GHz channel spacing -- we achieve a spectral efficiency of 4 bit/s/Hz and BER of all channels lower than the threshold of 20 \% hard-decision error correction (see Fig.\ref{fig:SI_MML}-\ref{fig:SI_MML_WDM_MDM_setup}). This setup showed a 1.28 Tb/s aggregated data rate using only 320 GHz of the spectral bandwidth. \\

\noindent\textbf{Chip-to-chip interconnect}  

\noindent The WDM-MDM communication link can be implemented for chip-to-chip interconnects onto which multiple chips and computing nodes are connected. In this section, we demonstrate a chip-to-chip MDM link using inverse-designed beam emitters and a rectangular core multimode fibre\cite{Marom:2017:IPC}, shown schematically in Fig.\ref{fig:Fig4}a. 

The chip-to-chip interconnect requires an efficient means of coupling light from on-chip transceivers to a communication link. In both free-space and fibre systems, this can be achieved using a beam emitter\cite{Yoo:2012:OE,Cai:2012:Science,Baumann:2018:OFC,Tsang:2019:JQE} that preserves and launches all spatial modes perpendicular to the chip. Here, the beam emitter structure is designed for an operation bandwidth of 40 nm with an 80-nm minimum feature size using photonic inverse design. The output of the MDM multiplexer is adiabatically tapered into the inverse-designed beam emitter to send data into orthogonal spatial modes and maintain the mode profiles from a silicon waveguide to free space or fibres. The dimensions of the beam emitter are flexible and can be varied over a wide range to maximize chip-to-fibre coupling efficiency. 

A modulated CW laser (NRZ at a symbol rate of 10 GBd) is first coupled to the Tx chip through one of four single-mode input waveguides via a lensed fibre. The Tx chip then sends light into a unique spatial mode of the multimode fibre using the MDM multiplexer (Fig.\ref{fig:Fig2}-\ref{fig:Fig3}) and inverse-designed grating (see Fig.\ref{fig:Fig4}b). The five-meter-long multimode fibre, which has a rectangular core (dimension: 32 $\times$ 8 $\mu$m$^{2}$; see the inset of Fig.\ref{fig:Fig4}a) constrained to support a single mode in the narrow transverse direction and four modes in the orthogonal direction, transmits the signal from the Tx chip to the Rx chip while maintaining low modal crosstalk. Finally, the Rx chip de-multiplexes the multimode signals and allows us to individually characterize the transmitted signals at single-mode output waveguides using a second lensed fibre.

Fig.\ref{fig:Fig4}c presents the measured channel cross-talks of the chip-to-chip MDM link using a multimode fibre. To benchmark the multimode fibre link, we also conduct a free-space experiment where the orthogonal spatial mode channels are transmitted between chips (see Fig.\ref{fig:SI_free_space} for experimental setup and detailed information). The cross-talk of the multimode fibre link ranges from -4.2 dB to -13.9 dB, and a 3-dB-bandwidth of chip-to-fibre coupling is approximately 60 nm for all mode channels (see Fig.\ref{fig:SI_MMF_coupling}). The mode-dependent insertion loss difference is approximately 8 dB at 1540 nm throughout the entire link. The variations are attributed to fibre-to-chip alignment and mode-dependent transmission loss of the fibre, which has higher loss in higher-order modes -- in fact, mode channel 1 - 3 show less than 2.5 dB insertion loss. Given that the cross-talk is dominated by the fibre-to-chip coupling alignment, the communication link quality can be greatly improved with appropriate photonic packaging\cite{IMEC:2017:OFC}. Fig.\ref{fig:Fig4}d shows eye diagrams of the chip-to-chip link using a multimode fibre link, driven with NRZ at a symbol rate of 10 Gb/s.\\

\noindent{\bf Conclusion}

\noindent In summary, we have demonstrated a multi-dimesional optical communication scheme using silicon photonic circuits. With an inverse-designed silicon MDM device and chip-scale soliton microcombs, we multiplex four spatial modes and{\color{red}} comb channels covering the entire C-band to achieve natively error-free ($<$ 10$^{-12}$) data transmission in 42 out of 52 channels through a nanophotonic waveguide. Additionally, we show chip-to-chip multi-mode communications enabled by inverse-designed beam emitters and a rectangular core multimode fibre. The MDM technology is fully compatible with silicon photonic WDM transmitters and semiconductor foundry processes. Here, photonic inverse design ensures device compactness and broadband operation. This enhances robustness to fabrication errors and temperature fluctuations, enabling practical applications in data centers and high performance computing. 

Scaling of this technology for ultra-wide bandwidth high-fidelity communication will require the maintenance of low mode cross-talk and insertion loss. In particular, we observe signal degradation in chip-to-chip data transmission experiments (both via multimode fibre and free-space). This signal degradation can be compensated by using MZI meshes\cite{Miller:2017:LSA} or digital MIMO signal processing\cite{Khan:2016:OL,Willner:2015:ScientificReport,Bell:2020}. Future scaled-up implementations of our communication scheme can employ such error-correction optical circuits on the receiver chip, and their performance can be improved in terms of footprint, insertion loss and operation bandwidth using photonic inverse design. Furthermore, efficient edge couplers\cite{Lipson:2020:CLEO,Shen:2020:OE}, foundry-compatible bi-level gratings\cite{Vuckovic:2018:OE,Popovic:2016:OFC} as well as advanced fibre technologies\cite{Puttnam:2021:Optica} can further mitigate mode cross-talk and insertion loss. 

As our MDM device features non-resonant, low-loss, low-crosstalk operation over the entire C-band, it is compatible with a large number of novel light sources, not limited to a Si$_{3}$N$_{4}$-based soliton microcomb (Fig.\ref{fig:Fig3}) and a table-top mode-locked laser (Fig.\ref{fig:SI_MML}) demonstrated in this work. For example, different types of chip-scale lasers and frequency comb sources such as mode-locked quantum dot lasers\cite{Bowers:2019:Optica}, electro-optical frequency combs\cite{Loncar:2019:Nature,Idjadi:2020:CLEO}, and vertical cavity surface emitting laser (VCSEL)\cite{Bell:2021:ECOC} can be used. Co-design of the laser source, optical link architecture, photonic devices and signal processing will facilitate the next generation of optical interconnects in data-centre networks, wireless communications, and hardware accelerators. \\

      

\bibliography{Reference}

\clearpage

\noindent{\bf Methods}\\
\noindent\textbf{Photonic inverse design} Stanford Photonics Inverse Design Software (SPINS)\cite{SPINs,Su:2020} was used to design mode-division multiplexers and surface-normal grating couplers for TE-polarized light. The inverse design method can provide a variety of photonic circuit component designs\cite{Wu:2019:LPR,Liu:2019:NatureCommunications,Gabrielli:2020:PTL,Piggott:2020:ACSPhotonics} (see Fig.\ref{fig:SI_MMBS}) as well as a potential compact 12 channel multiplexer design (see Fig.\ref{fig:SI_12by12}).  \\

\noindent\textbf{Optoelectronic transceiver} The AMF photonic process is a 180 nm silicon-on-insulator fabrication process with 2 $\mu$m buried oxide thickness. The thickness of the silicon layer is 220 nm and the single-mode (500 nm width) photonic waveguides have propagation loss of about 2 dB/cm. The grating coupler is designed by partially etching silicon and has a measured loss of about 4 dB\cite{Ashtiani:2019:Optica,Idjadi:2020:NaturePhotonics}. Also, the Y-junction has a measured excess loss of 0.5 dB\cite{Zhang:2013:OE}. The implemented PN ring modulator has a free spectral range of about 11 nm and a 3-dB bandwidth of higher than 30 GHz\cite{Ashtiani:2019:Optica}. The photodiode has a responsivity of higher than 0.75 A/W and a 3-dB bandwidth of more than 30 GHz\cite{Ashtiani:2019:Optica,Idjadi:2020:NaturePhotonics}. Fig.\ref{fig:SI_IME_eye_diagram} shows measured results of data transmission from electro-optic ring modulator to MDM components and Si/Ge photodiode. In addition, we show the channel cross-talk of the all-passive MDM device.  \\

\noindent\textbf{Soliton-WDM-MDM data transmission} The soliton microcombs are generated by pumping the silicon nitride microresonator with a CW laser, IQ modulator in a single-sideband suppressed-carrier configuration driven by a voltage-controlled oscillator, and a subsequent amplifier; see Fig. \ref{fig:SI_soliton_WDM_MDM_setup} and Ref\cite{Papp:2020:PhyRevAppl} for a more detailed description of the comb generation setup and silicon nitride device designs. The output of the comb is then passed through a WDM filter that serves to suppress the pump power, and the transmitted comb power is amplified and de-multiplexed using a commercial 100 GHz ITU grid based DWDM with an insertion loss of 3.5 dB per channel. The `even' and `odd' carriers are recombined using another multiplexer. Each channel is separately amplified and passed through two intensity modulators (20 GHz 3-dB bandwidth and 6-dB insertion loss, EOSpace) which are driven by PRBS generators using non-return to zero (at a data rate of 10 Gb/s). The data channels are de-correlated with a delay of approximately 5000 symbols. We recombine the odd and even sets of carriers, amplify them using an EDFA and split the power into four different channels. Mode channel 2 - 4 have a delay of 50, 100 and 200 symbols with respect to mode channel 1. Output coupling is performed with a lensed fibre aligned to one MDM output waveguide at a time. The light at the output is then amplified and sent through another de-multiplexer where the signal is detected using a 14 GHz bandwidth photodiode (DSC40S) with a receiver responsivity of 0.8 A/W. The photodetected signal is then sent to an error analyzer which we use to measure the bit-error rate. Fig. \ref{fig:SI_soliton_WDM_MDM_setup} shows a detailed experimental setup.\\

\noindent\textbf{MLL-WDM-MDM data transmission} A mode-locked laser (Lumentum, repetition rate: 10-GHz) generates a frequency comb. The frequency comb is passed through a delay line interferometer (free-spectral-range: 20-GHz) to increase the optical comb spacing. Then the 20-GHz comb is passed through a highly nonlinear fibre (HNLF) for spectral flattening and broadening\cite{Willner:2014:OE}. The frequency lines of the optical comb are selectively filtered and de-interleaved into `odd' and `even' carriers by a programmable filter (Finisar WaveShaper). The `odd' and `even' carriers are separately amplified using C-band EDFAs, and routed to in-phase/quadrature modulators which are driven by arbitrary-waveform generators using quadrature phase-shift keying (at a symbol rate of 10-Gbd). The data channels are decorrelated by 50 symbols delays and combined into a standard single-mode fibre. Next we boost the power of all of the data channels using an EDFA and simultaneously launch them into the four input ports of the MDM chip using a fibre array (single-mode fibres with 127 $\mu$m spacing). Output coupling is performed with a tapered lensed fibre aligned to one MDM output waveguide at a time, and the received signal is characterized using intradyne coherent detection with electrical filtering and a tunable external cavity laser as a local oscillator. Fig.\ref{fig:SI_MML_WDM_MDM_setup} shows the experimental setup and optical spectra of MLL sources.          \\

\noindent\textbf{Four spatial mode rectangular core fibre} An optical fibre with a rectangular core geometry was provided by Corning Incorporated. The germanium-doped core material was machined to rectangular shape and with an annular silica cladding material was drawn down to final dimensions (125 $\mu$m cladding diameter, nearly perfectly circular, with core of 32 $\times$ 8 $\mu$m$^{2}$ and index modulation $\sim$ 0.005). The fiber supports four polarization-degenerate spatial modes with intermode effective index differences $\ge$ 0.0005. A 5-meter-long fibre sample was used to image the four guided modes using low-coherence interferogram image analysis (beating against a plane wave, see Fig.\ref{fig:SI_MMF}). Measured differential modal group delay was $\sim$ 8 ns/km, which can lead to signal skew in long mode multiplexed links and MIMO processing complexity if mode mixing occurs.  \\

\noindent\textbf{Free-space communication experiments} Optical signals were generated using a CW tunable laser, passed through a circulator, and modulated with an amplitude electro-optic modulator. The modulator was driven by a data generated by micro-controller and electronic amplifier. To recover from modulator losses and to compensate for extraneous losses in the free space system, the signal was amplified with a fibre amplifier before coupling to the transmitter chip. The amplified beam passes through a polarization controller, free space coupler, and 50/50 beamsplitter and is focused onto input grating couplers through a 100$\times$ nIR objective lens (Mitutoyo). Light that is coupled out of the multimode output grating passes through the beamsplitter a second time and is collimated with a 200mm focal length lens. This beam is sent approximately 0.5 m to another beamsplitter, 200 mm lens, and 100$\times$ objective that focus the multimode beam onto the multimode grating of the Rx chip. The output of the Rx chip is coupled out once more from the 100$\times$ objective and coupled to a single-mode fibre. 10\% of the light from the laser is picked off with a directional coupler, passed through a polarization controller, and 1\% of this light is coupled to the output fibre. The homodyne signal is read using a photodiode and a custom electric amplifier chain fed to an oscilloscope. Fig.\ref{fig:SI_free_space} shows the experimental setup, free-space data transmission, and measured optical mode profiles at multiple wavelengths. \\

\noindent\textbf{Data availability} The data that support the plots within this paper and other findings of this study are available from the corresponding author upon reasonable request.\\

\noindent\textbf{Code availability} An open source of the photonic optimization software used in this paper is available at https://github.com/stanfordnqp/spins-b.\\

\medskip

\noindent\textbf{Acknowledgment}
\noindent We acknowledge insightful discussion with J.M.Kahn, S.Fan, A.Dutt, and H.Kwon, and are also grateful for technical advice from S.K.Pai, M.H.Idjadi, and D.Huang. The chip-scale devices were fabricated in the Stanford Nanofabrication Facility, the Stanford Nano Shared Facilities, Advanced Micro Foundry Pte Ltd, AIM photonics, and LIGENTEC. This work is funded by the DARPA under the PIPES program, and the AFOSR under the MURI program (Award No. FA9550-17-1-0002). A.E.W. acknowledges the support from from AFRL (FA8650-20-C-1105). We thank G.Keeler, G.Pomrenke, and the programme management teams for discussions throughout the project. \\

\noindent\textbf{Competing interests.}
The authors declare they have no competing financial interests.

\clearpage

\renewcommand{\thefigure}{S\arabic{figure}}
\setcounter{figure}{0}

\begin{figure*}[t!]
\centering
\includegraphics[width=\linewidth]{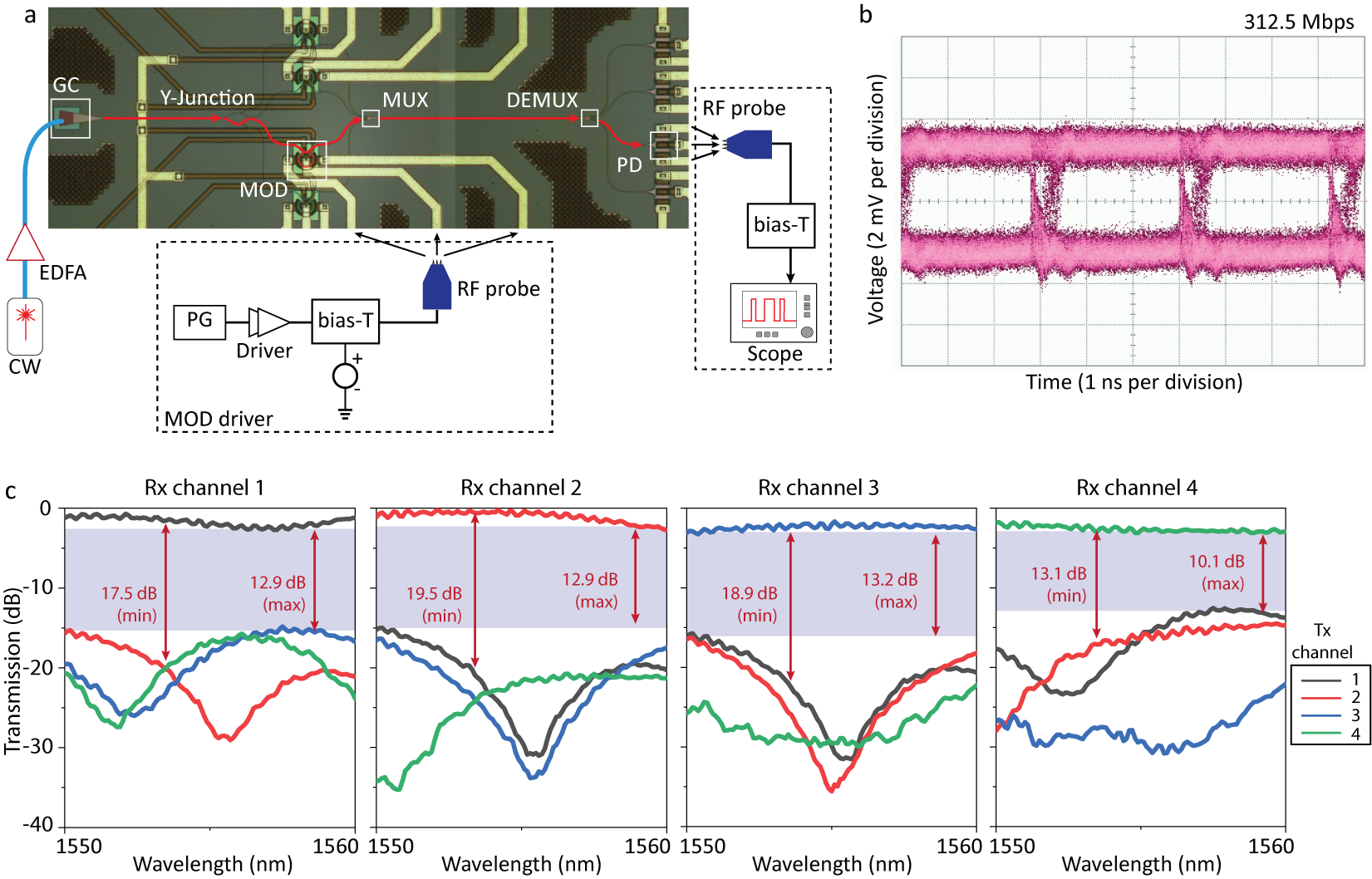}
\caption{\label{fig:SI_IME_eye_diagram}{\bf{Integrated MDM transceiver.}} (\textbf{a}) Data transmission measurement. A CW laser is coupled to an integrated MDM transceiver (GC: grating coupler, MOD: ring modulator, PG: pattern generator, MUX/DEMUX: MDM multiplexer/demultiplexer, PD: photodiode). The cascaded Y-junctions evenly distribute optical power and simultaneously send a light to four ring modulators -- one of ring modulators is driven by an external electric driver. We measured transmitted data using integrated Si/Ge photodiodes (PD) and an external scope. (\textbf{b}) Measured eye diagram of MDM channel 3 (TE$_{20}$) obtained from an integrated Si/Ge photodiode (PD) and external signal scope. (\textbf{c}) Measured S-parameters of the back-to-back MDM multiplexer structure. We sent the multiplexer design in (\textbf{a}) to two semiconductor foundries (AMF and AIM photonics), and channel cross-talks at 1550 - 1560 nm (ring modulator operation wavelength) of all channels consistently ranges from -10 dB to -20 dB on chips from two foundries. }
\end{figure*}

\begin{figure*}[t!]
\centering
\includegraphics[width=0.8\linewidth]{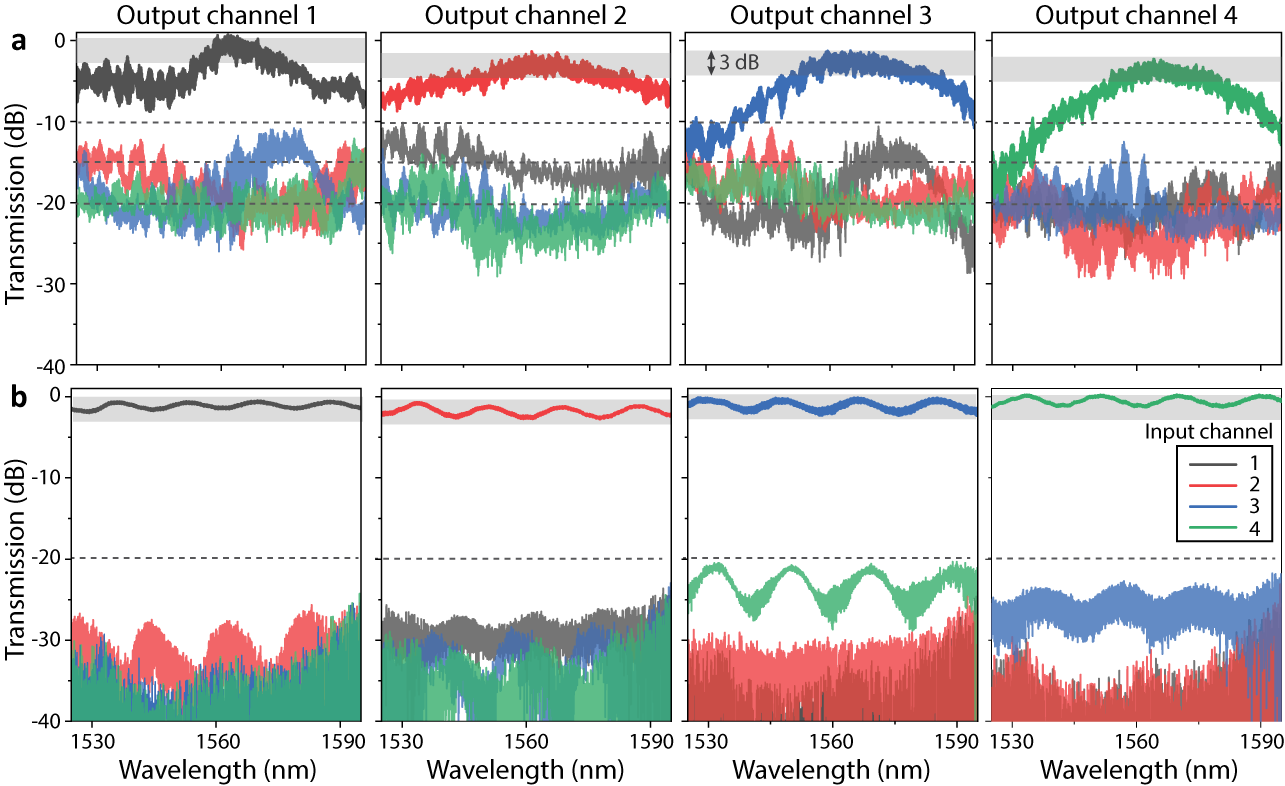}
\caption{\label{fig:SI_MDM_design}{\bf{Optimization of broadband, low-crosstalk MDM multiplexer.}} Measured S-parameters of the back-to-back MDM multiplexer structures with different design bandwidths. The structure in (\textbf{b}) is optimized at multiple wavelengths over the entire C band, while the structure in (\textbf{a}) is optimized only at a single wavelength (1560 nm). The narrow band structure is implemented in the WDM-MDM communications using a mode-locked laser (16 wavelength channels over 320 GHz spectral span; see Fig.\ref{fig:Fig3}), and the broadband structure is utilized in the WDM-MDM communications using soliton microcombs (13 wavelength channels over 3.6 THz spectral span; see Fig.\ref{fig:Fig3}).}
\end{figure*}

\begin{figure*}[t!]
\centering
\includegraphics[width=0.95\linewidth]{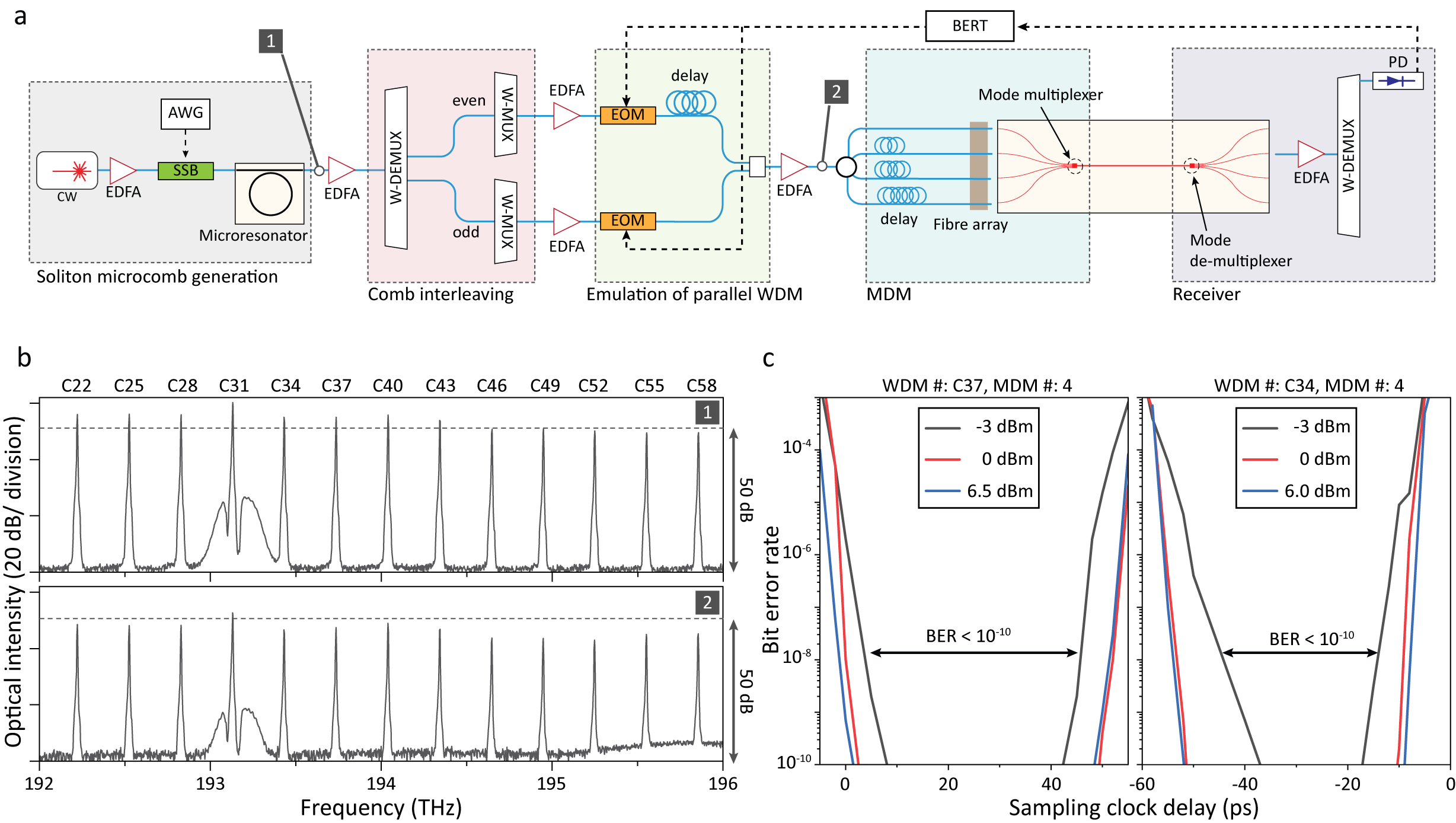}
\caption{\label{fig:SI_soliton_WDM_MDM_setup}{\bf{Soliton microcomb data transmission.}} \textbf{(a)} A silicon nitride microresonator generates a frequency comb with a 300 GHz repetition rate (optical spectrum: top panel of \textbf{(b)}). The comb lines are de-multiplexed using a standard ITU grid DWDM (W-DEMUX), and de-interleaved into `odd' and `even' carriers using another multiplexer (W-MUX). The `odd' and `even' channels are amplified and passed through intensity modulators (EOM) which are driven by PRBS generator using a NRZ encoding. The data channels are decorrelated by a fibre delay and re-combined into a standard single-mode fibre (optical spectrum: bottom panel of \textbf{(b)}). The combined sets of carriers are amplified and simultaneously coupled to four single-mode waveguide inputs of the silicon MDM chip (see Fig.\ref{fig:Fig2}c), and the transmitted light through the MDM chip is coupled back to single-mode fibre using a lensed fibre. The collected light is amplified and sent through another demultiplexer (W-DEMUX) where the received signal is analyzed using a photodiode (PD) and an error analyzer. \textbf{(b)} Frequency comb spectra before (top panel) and after (bottom panel) optical amplification. \textbf{(c)} 10-Gb/s bathtub curves obtained from BERT by sweeping the delay between the clocks for transmitter and receiver. We measured a bit-error rate better than 10$^{-10}$ with -3 dBm received optical power. Left: Measured curves of the channel C37 (193.7 THz)/ M4 (TE$_{30}$), Right: Bathtub curves of the channel C34 (193.4 THz)/ M4 (TE$_{30}$) at various received optical powers.}
\end{figure*}

\begin{figure*}[t!]
\centering
\includegraphics[width=0.75\linewidth]{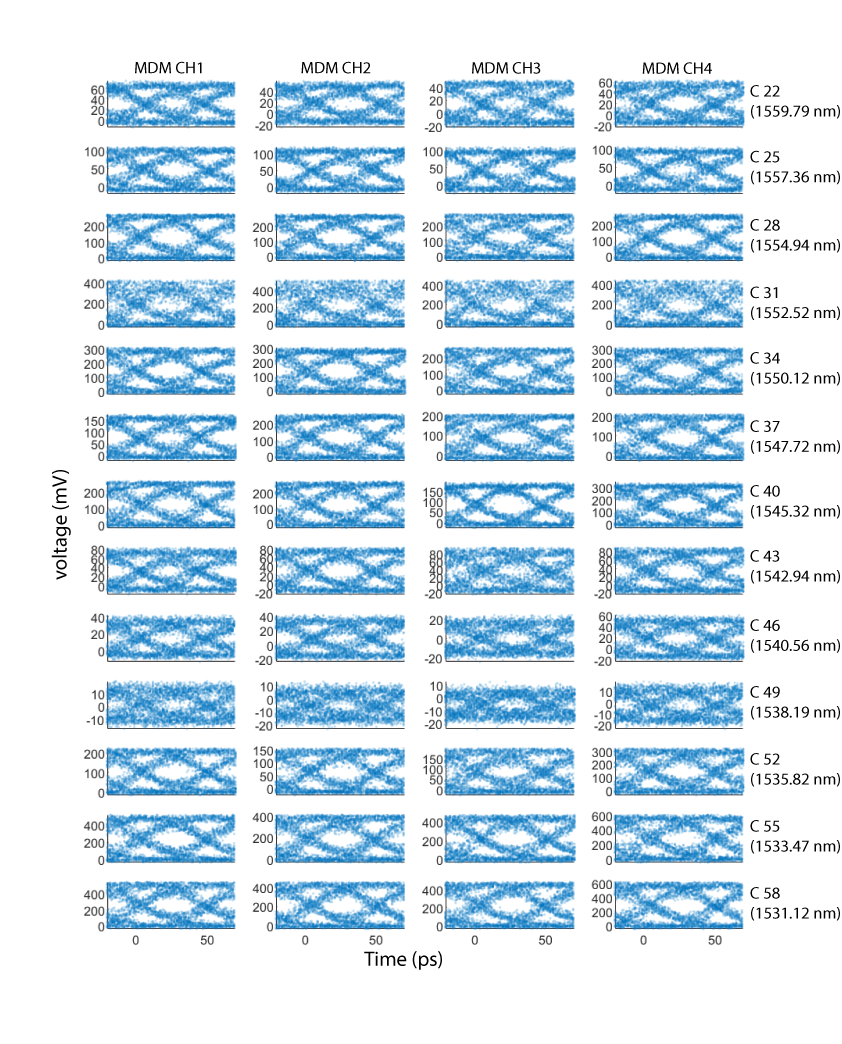}
\caption{\label{fig:SI_soliton_WDM_MDM_20gbps_eye_diagram}{\bf{Eye diagrams of all data channels (13 wavelength channels $\times$ 4 mode channels) at 20 Gb/s NRZ.}} }
\end{figure*}

\begin{figure*}[t!]
\centering
\includegraphics[width=\linewidth]{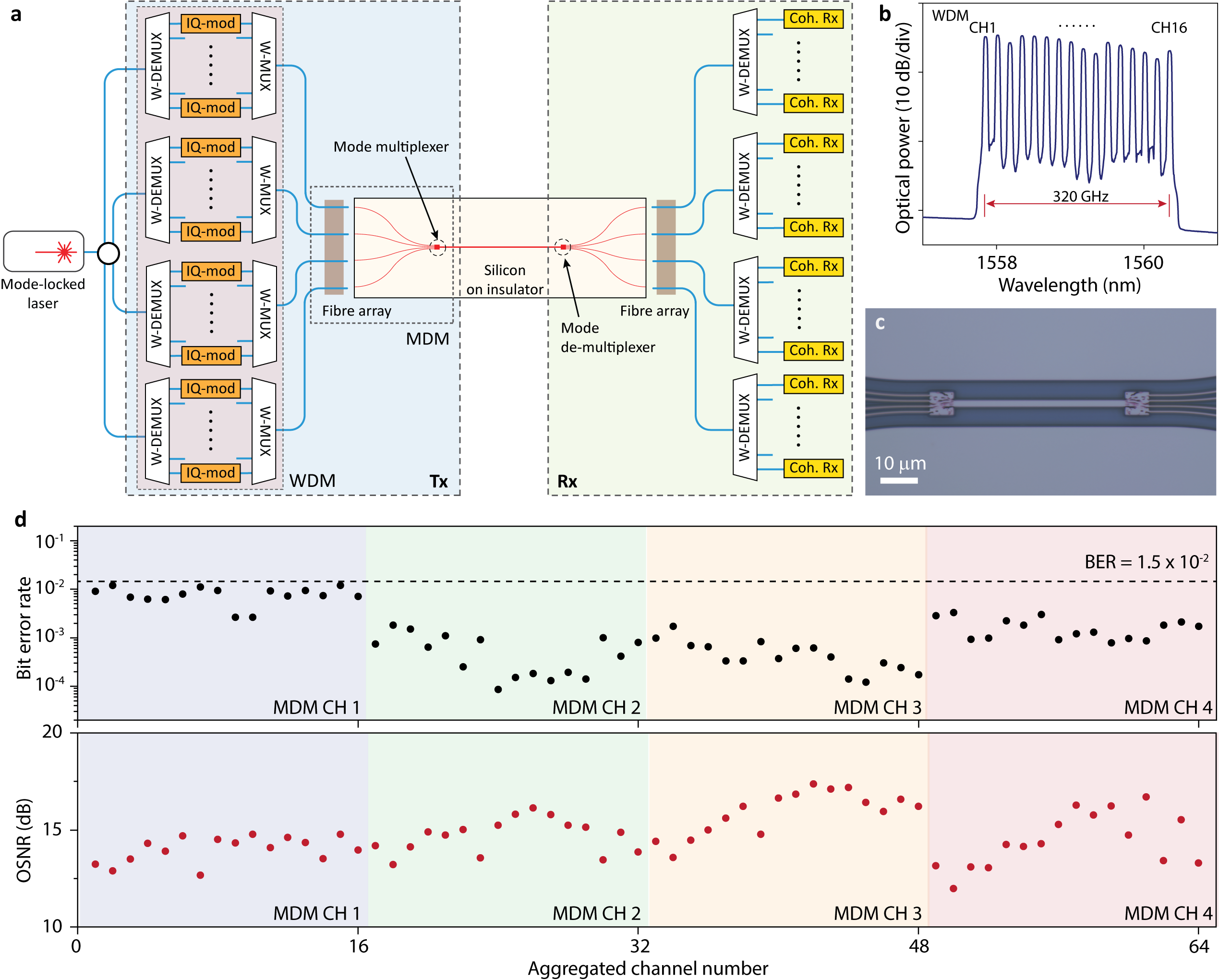}
\caption{\label{fig:SI_MML}{\bf{On-chip interconnect: WDM - MDM data transmission with a frequency comb laser.}} \textbf{(a)} A mode-locked laser is employed as a multi-wavelength laser source and evenly distributed to four WDM transmitters. WDM de-multiplxers (W-DEMUX) separate the comb lines (16 WDM channels) and in-phase/quadrature modulators (IQ-mod) encode independent data (QPSK at a symbol rate of 10 GBd per WDM channel). The WDM data are recombined into single-mode fibres using WDM multiplexers (W-MUX), and four fibres are coupled to MDM input ports simultaneously using a fibre array. The on-chip mode multiplexer recombines all data channels (aggregated channel numbers: 4 $\times$ 16) into a multimode waveguide, and sends WDM-MDM data to the receiver. At the receiver, the mode and wavelength channels are separated by mode and wavelength de-multiplexers and detected using digital coherent receivers (Coh. Rx). In our experiments, we independently modulate even and odd WDM channels using two IQ-modulators for emulating WDM transmission (see Fig.\ref{fig:SI_MML_WDM_MDM_setup} for more details). \textbf{(b)} Optical spectrum of the frequency comb laser (see method section for more details). \textbf{(c)} Optical microscope image of MDM multiplexer back-to-back structure. \textbf{(d)} Top: measured BERs of the transmitted data channels along with BER threshold when applying forward error correction with 20 \% overheads (dashed black line), Bottom: Optical signal-to-noise ratio (OSNR) of the transmitted channels.  }
\end{figure*}

\begin{figure*}[t!]
\centering
\includegraphics[width=0.95\linewidth]{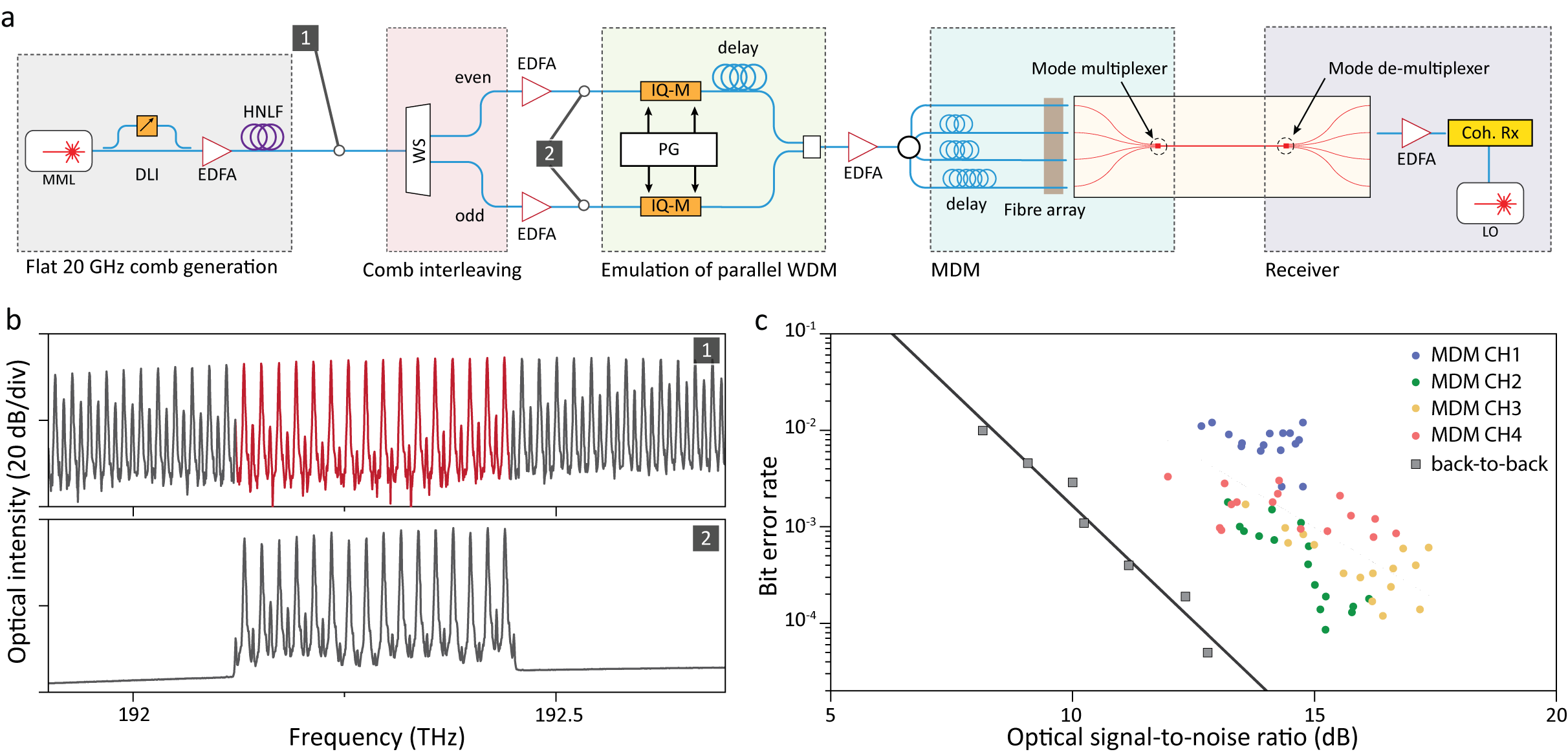}
\caption{\label{fig:SI_MML_WDM_MDM_setup}{\bf{Data transmission using flattened mode-locked laser for parallel WDM and MDM.}} \textbf{(a)} A mode-locked laser generates a frequency comb. The frequency comb is passed through a delay line interferometer (DLI) to increase the optical comb spacing. Then the 20-GHz comb is passed through a highly nonlinear fibre (HNLF) for spectral flattening and broadening. Frequency combs (optical spectrum: top panel of \textbf{(b)}) are selectively filtered and de-interleaved into `odd' and `even' carriers using a waveshaper (WS). Each carriers are separately amplified (optical spectrum: bottom panel of \textbf{(b)}) and passed through IQ modulators (IQ-M) which are driven by arbitrary-waveform generators using QPSK. The data channels are decorrelated by a fibre delay and combined into a single-mode fibre. All carriers are amplified and simultaneously coupled to four single-mode waveguide inputs of the silicon MDM chip (see Fig.\ref{fig:Fig2}b). We collect light at the waveguide output using a lensed fibre, and the collected light is analyzed using a coherent receiver and local oscillator (LO). (\textbf{b}) Measured optical spectra of flattened 20-GHz frequency comb (top) and selectively filtered comb tooth after interleaving and amplification (bottom). (\textbf{c}) Measured BER versus optical signal-to-noise ratio of all 64 channels as well as back-to-back test case.   }
\end{figure*}

\begin{figure*}[t!]
\centering
\includegraphics[width=0.9\linewidth]{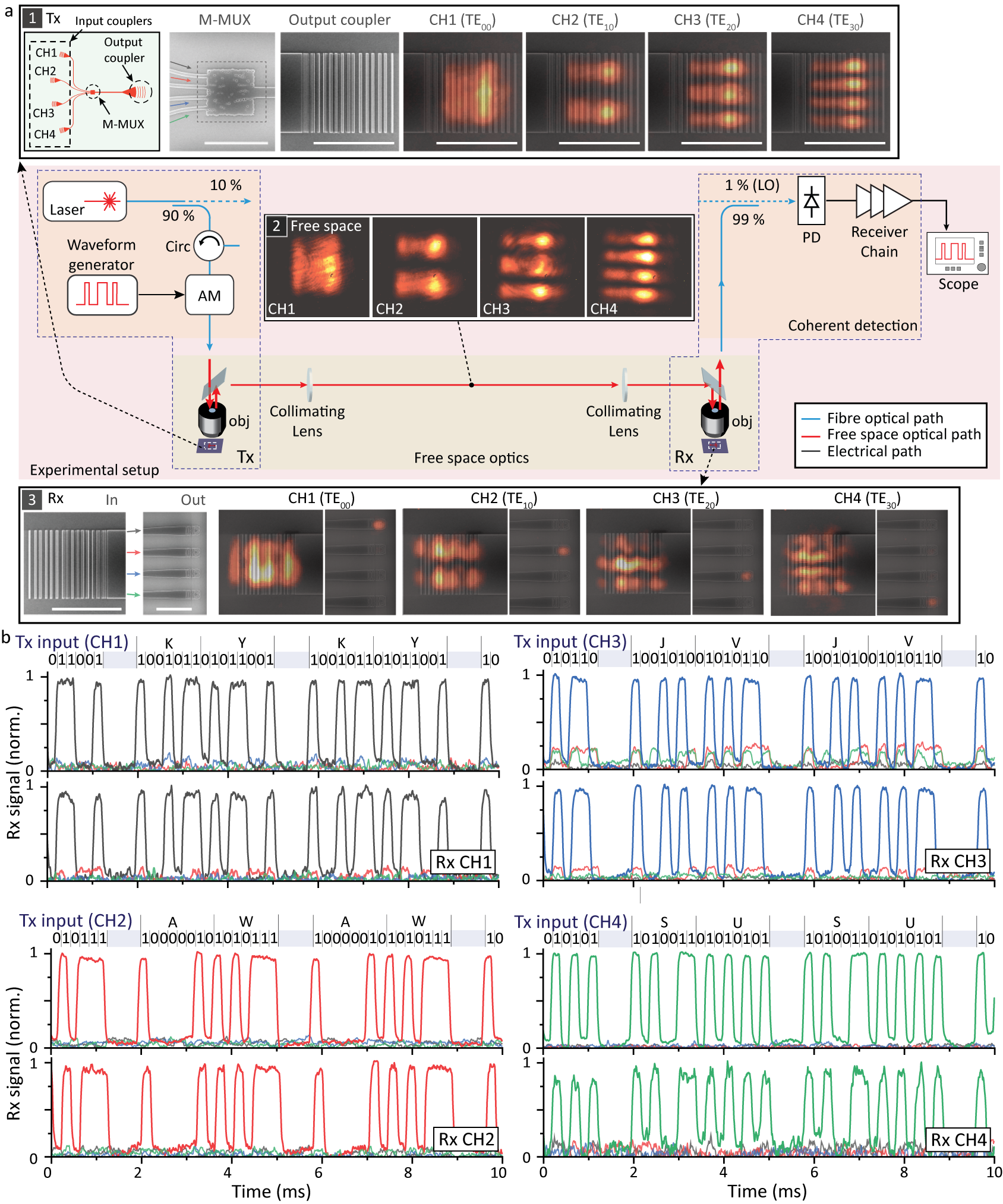}
\caption{\label{fig:SI_free_space}{{\bf{Chip-to-chip interconnect: Wireless data transmission using inverse-designed beam emitter.}} \textbf{(a)} A CW laser is modulated with on-off-keying data using an amplitude modulator, and this data signal is coupled to the Tx input grating couplers. The Tx chip emits multi-channel MDM data to free-space using an inverse-designed coupler (Inset 1: measured optical mode profile at Tx grating coupler, Inset 2: mode profile in free-space between Tx and Rx chips). The free-space modes are projected onto an MDM demultiplexing device in the Rx chip that separates each mode into a single-mode output (Inset 3). The transmitted data are detected using a coherent receiver with the CW laser as the local oscillator (LO). Infrared images are overlaid on device SEM images (bar: 10 $\mu$m), and measured at 1525 nm. \textbf{(b)} Received data traces at each Rx output grating coupler (Top: 1525 nm, Bottom: 1540 nm). The signal cross-talk at Rx CH3 trace is attributed to the effects of channel-dependent optical link insertion loss, the limited size aperture, and misalignment.}}
\end{figure*}

\begin{figure*}[t!]
\centering
\includegraphics[width=0.6\linewidth]{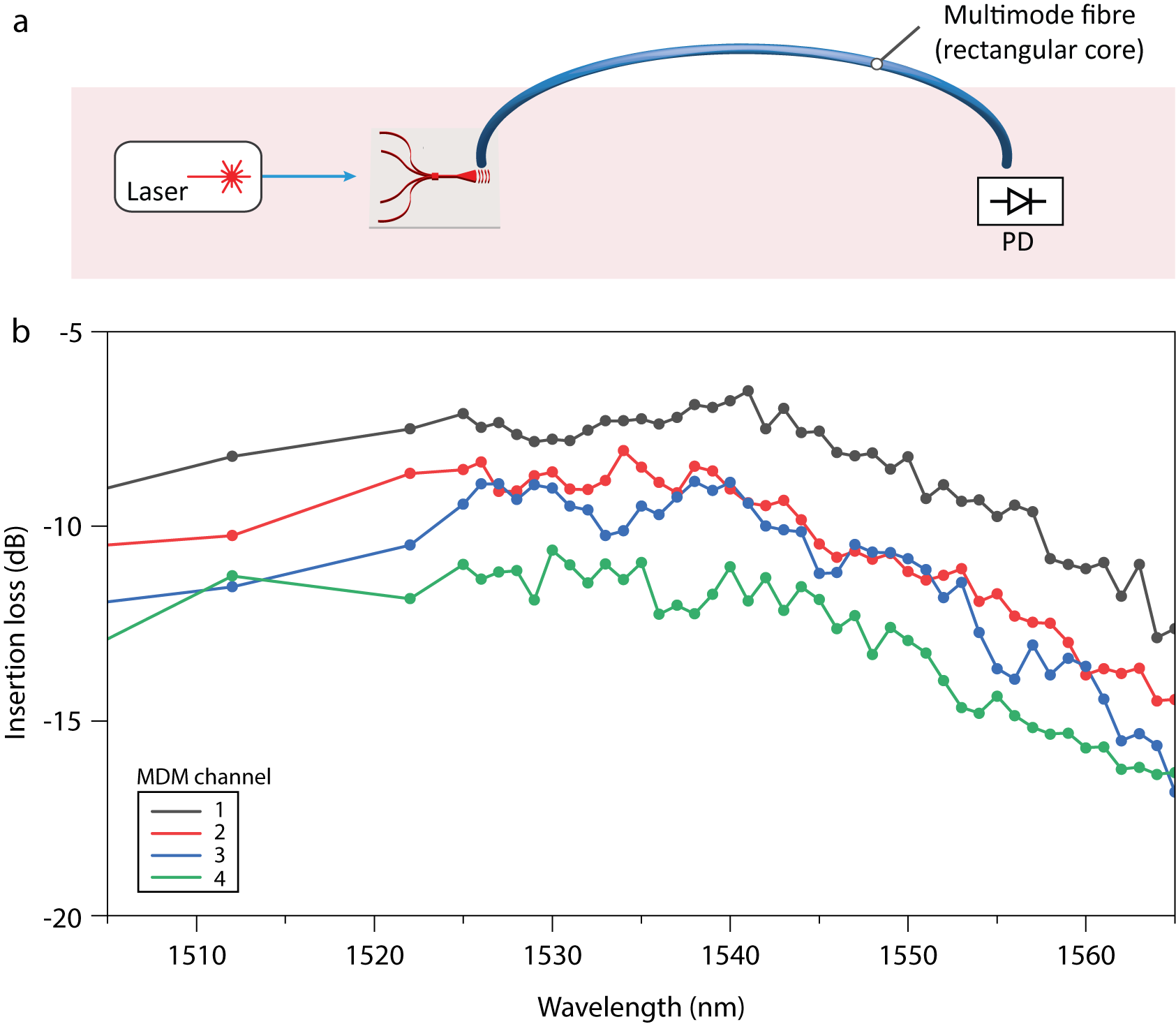}
\caption{\label{fig:SI_MMF_coupling}{{\bf{Mode-dependent loss measurement.}} \textbf{(a)} A CW laser is coupled to the MDM chip, and we measured transmitted optical power through MDM multiplexer, grating coupler, and a ractangular core multimode fibre using a photodetector.  \textbf{(b)} Measured insertion loss of mode channels exhibits a 3-dB-bandwidth of 60 nm for all channels, and the mode-dependent loss difference is approximately 5 dB. The mode-dependent loss is attributed to fibre transmission loss and chip-to-fibre misalignment. }}
\end{figure*}

\begin{figure*}[t!]
\centering
\includegraphics[width=\linewidth]{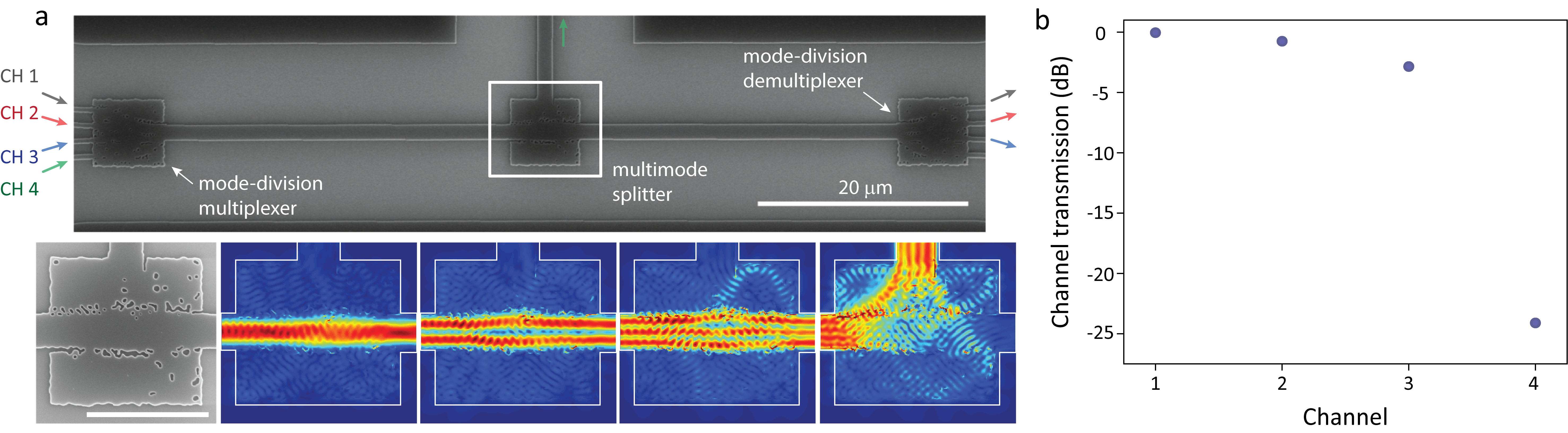}
\caption{\label{fig:SI_MMBS}{\bf{Arbitrarily routed multimode photonic circuit element.}} Inverse design can provide a multimode photonics toolkit. We combine functionalities of a multimode crossing and waveguide bend\cite{Wu:2019:LPR} in a compact design area. \textbf{(a)} Top: SEM image of a multimode photonic circuit which consists of an MDM multiplexer, multimode splitter, and MDM demultiplexer. The multimode splitter (white rectangular box, design area: 6 $\mu$m $\times$ 6 $\mu$m) can selectively route multimode signals to different waveguides. Channels 1-3 are sent to the demultiplexer while the channel 4 is routed to the other waveguide port. Bottom: Zoomed-in SEM image (bar: 5 $\mu$m) and simulated multimode routing. \textbf{(b)} Measured transmission for the multiplexer-splitter-demultiplexer structure. }
\end{figure*}

\begin{figure*}[t!]
\centering
\includegraphics[width=\linewidth]{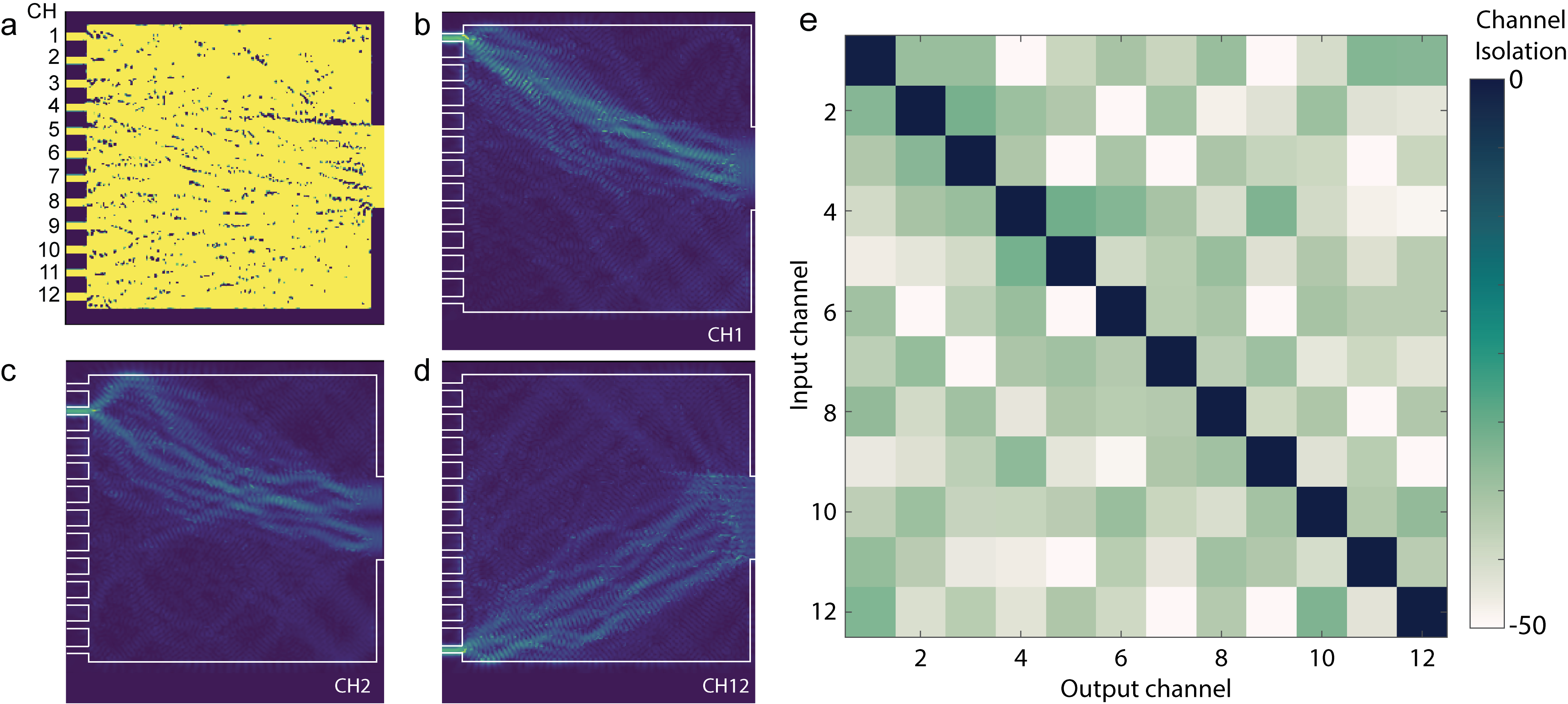}
\caption{\label{fig:SI_12by12}{\bf{Optimization of 12 channel mode-division multiplexer.}} For computational simplicity, we restrict our study to a two-dimensional structure with transverse-electric polarization. We model the structure with infinite extent in the third dimension and an effective relative permittivity of a silicon slab (220-nm-thick silicon at 1550 nm wavelength). \textbf{(a)} Structure of optimized device (18 $\mu$m $\times$ 18 $\mu$m), which consists of twelve single-mode input waveguides and one multi-mode output waveguide \textbf{(b-d)} Mode 1-, 2- and 12- transmission through device. \textbf{(e)} S-parameters of simulated device (note that it is a single multiplexer device and it is not a simulation of back-to-back structure). The channel cross-talk ranges from -53 dB to -25 dB, and insertion loss ranges from from -0.23 dB to -0.47 dB.}
\end{figure*}

\begin{figure*}[t!]
\centering
\includegraphics[width=0.7\linewidth]{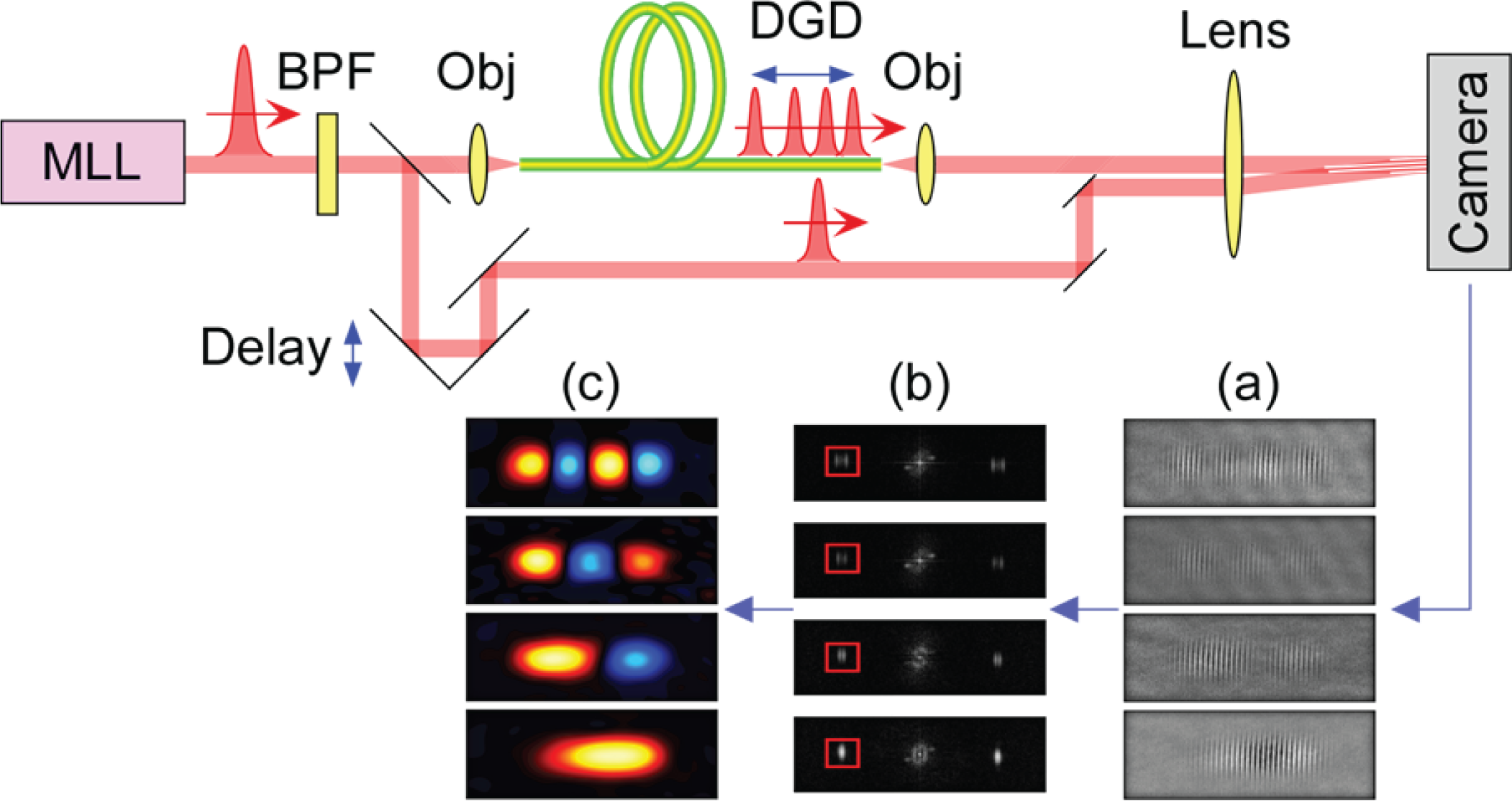}
\caption{\label{fig:SI_MMF}{\bf{Four spatial mode rectangular core fiber.}} Experimental setup for imaging fiber modes using low-coherence interferogram analysis. Short optical pulse from a mode-locked laser at center wavelength of 1.55 $\mu$m is focused with an objective (Obj) into the rectangular core fiber and excites all its spatial modes. (An optional band-pass filter (BPF) is used to increase the coherence lengths). At the fibre’s distal end, the modes acquire differential group delays (DGD). A copy of the excitation pulse is timed for co-incidence onto an infrared camera, obtaining an interferogram of one mode at a time, inset (a). The mode profiles are extracted following a Fourier transform and selection of the off-axis diffracted signal (inset (b)), and inverse Fourier transform to reveal the mode profile, inset (c).}
\end{figure*}

\end{document}